\documentclass[aps,prd,twocolumn,floats,superscriptaddress,nofootinbib]{revtex4-1}

\usepackage{graphicx}
\usepackage{dcolumn}
\usepackage{bm}
\usepackage{parskip}
\usepackage{amsmath,amssymb}
\usepackage{indentfirst}
\usepackage{slashed}
\usepackage{mathtools}
\usepackage{braket}
\usepackage{color}
\usepackage{commath}
\usepackage{subfigure}

\begin{document}


\title{A variant two-Higgs doublet model with a new Abelian gauge symmetry}

\author{Chuan-Ren Chen}
 \email{crchen@ntnu.edu.tw}
\affiliation{Department of Physics, National Taiwan Normal University, Taipei, Taiwan 11677, R.O.C.}%

\author{Cheng-Wei Chiang}%
 \email{chengwei@phys.ntu.edu.tw}
\affiliation{Department of Physics, National Taiwan University, Taipei, Taiwan 10617, R.O.C.}
\affiliation{Institute of Physics, Academia Sinica, Taipei, Taiwan 11529, R.O.C.}
\affiliation{Department of Physics and Center of High Energy and High Field Physics, National Central University, Chungli, Taiwan 32001, R.O.C.}

\author{Kuo-Yen Lin}%
 \email{r06222069@ntu.edu.tw}
\affiliation{Department of Physics, National Taiwan University, Taipei, Taiwan 10617, R.O.C.}

\date{\today}

\begin{abstract}
We consider a two-Higgs doublet model extended with a broken Abelian gauge symmetry under which all Standard Model (SM) quarks, fourth generation fermions and a new SM-singlet scalar boson are charged.  Such a setup is shown to be able to accommodate the muon anomalous magnetic dipole moment while being consistent with existing constraints of flavor-violating decays of charged leptons and $Z$ boson.  The new scalar boson offers a suitable dark matter candidate that interacts with the SM particles via the Higgs portal and the $Z'$ boson associated with the new gauge symmetry.  The dark matter direct detection bound is found to impose a strong constraint on the new gauge coupling.
\end{abstract}

\maketitle

\setlength\parindent{15pt} 
\setlength{\parskip}{10pt} 

\section{Introduction}
\label{sec:intro}

Despite its great success in explaining almost all the existing observational data, the Standard Model (SM) is widely believed to be just an effective theory of a more fundamental theory.  Even with the constraints of known symmetries and experimental data, there are still a plethora possible ways to extend the SM.  Particularly interesting ones are those that can accommodate some of the outstanding anomalies and/or address certain theoretical issues.

One trivial extension of the SM is to replicate another generation of fermions, the sequential fourth generation.  Besides its simplicity, the existence of fourth generation fermions may address some unanswered questions to the SM, such as electroweak symmetry breaking and bayon-antibaryon asymmetry~\cite{Holdom:1986rn,Hill:1990ge,Hung:2010xh,Hou:2011df,Fok:2008yg}.
However, this possibility is excluded due to the conflict between the mass lower bound  ($ \gtrsim 680$ GeV) given by the LHC direct search for sequential fourth generation quarks~\cite{Chatrchyan:2012fp} and the mass upper bound  ($\lesssim 550~{\rm GeV}$) required by the unitarity of the partial wave amplitude in fourth-generation top quark pair ($t^\prime \overline{t^\prime}$) scattering~\cite{Erler:2010sk}.  Furthermore, the existence of sequential fourth generation fermions is also ruled out by the current Higgs boson data at the LHC, as an enhancement by a factor of $9$ or so in the Higgs boson production via gluon-gluon fusion (ggF) is expected due to the additional contributions of fourth-generation top and bottom quarks ($b'$).

It has been recently shown that in the Type-II two-Higgs doublet model (2HDM) with sequential fourth generation fermions, both the Higgs data and the direct search at the LHC can be reconciled provided a subtle cancellation occurs in the Higgs ggF production~\cite{Das:2017mnu}.  In certain parameter space, dubbed the \emph{wrong-sign} scenario, the Yukawa coupling of $b'$ can have an opposite sign to its up-type partner $t'$ while they are almost degenerate in mass.  As a result, their contributions to the Higgs ggF production amplitude cancel with each other.

Anomalous magnetic dipole moments of electron and muon have been provided as tests for the SM, as they can be  both experimentally measured and theoretical computed to an extremely high precision.  Interestingly, the muon anomalous magnetic dipole moment, $a_\mu \equiv (g-2)_\mu/2$, has been observed to be $\sim 3\sigma$ away from the SM prediction since its first measurement.  This longstanding discrepancy, known as the muon $g-2$ anomaly, triggers many proposals beyond the SM (see, {\it e.g.}, Refs.~\cite{Jegerlehner:2009ry} and \cite{Blum:2013xva} for reviews).

Currently the experimental result shows that the discrepancy $\Delta a_\mu \equiv a_\mu^{\text{exp}} -  a_\mu^{\text{SM}} = (27.05 \pm 7.26) \times 10^{-10}$ reaches about $3.7\sigma$~\cite{Keshavarzi:2018mgv}.  With new data from the Fermilab experiment E989 in the near future, the significance is expected to be over $5\sigma$, assuming the central values of both theory and experiment results remain unchanged~\cite{Blum:2013xva}. 
If this is the case, the muon $g-2$ anomaly certainly involves contributions from beyond the SM.
In Type-II 2HDM, additional contributions to $a_\mu$ have been calculated up to two loops, and the result is not large enough to explain the difference, even with the contributions of sequential fourth generation fermions~\cite{Huo:2003pw,Cheung:2003pw,Broggio:2014mna}.

In this paper, we point out that with the introduction of a new scalar boson having lepton flavor-violating couplings, the one-loop diagram involving a heavy chiral lepton can be enhanced.  We will focus on the wrong-sign scenario in the type-II 2HDM containing fourth generation fermions, where the heavy charged chiral leptons are not canonically charged under the SM electroweak gauge group while the quarks stay sequential.  The new scalar $\phi$ is assumed to be an SM singlet.  It enhances the muon $g-2$ through its flavor-violating coupling between muon and the fourth-generation charged lepton $e_4$.

To avoid $\phi$ from having other undesirable couplings with fermions, we introduce a gauged $U(1)_X$ symmetry.  Under the new Abelian symmetry, all fermions except for the leptons in the first three generations have nonzero charges.  The two Higgs doublets are also neutral under $U(1)_X$. Assumed to be lighter than the fourth generation fermions, the $\phi$ field in the model offers a suitable dark matter (DM) candidate to explain the observed relic density.  Moreover, the DM direct search bound imposes a strong constraint on the gauge coupling associated with the new $U(1)_X$ symmetry.

The structure of this paper is organized as follows.  We start with the model setup in Section~\ref{sec:model}.  We present in Section~\ref{sec:gm2} how the main contribution from the heavy charged lepton fills the gap between the experimental data and SM prediction for the muon $g-2$.  We also show the constraints on the model parameter space due to the lepton flavor-violating decays of charged leptons and the $Z$ boson.  In Section~\ref{sec:dm}, we calculate the DM relic density and derive a bound from the direct search.  Finally, conclusions are given in Section~\ref{sec:conclude}.

\section{Model setup}
\label{sec:model}

We consider an extension of the usual type-II 2HDM with three right-handed neutrinos ($\nu_R$), a fourth generation of fermions:
\begin{align*}
\begin{split}
Q_{L4} \equiv \binom{u}{d}_{L4} ~,~ u_{R4} ~,~ d_{R4}
~,
\\
L_{R4} \equiv \binom{\nu}{e}_{R4} ~,~ e_{L4} ~,~ \nu_{L4}
~,
\end{split}
\end{align*}
and a SM singlet scalar $\phi$.  The model also has a new gauged $U(1)_X$ symmetry under which all but the leptons in the first three generations and the two Higgs doublets are charged.  As a result, the heavy fermions $t^\prime$, $b^\prime$ and $\ell_4$ have additional decay channels to SM particles and $\phi$, provided they are kinematically allowed.  In addition, as the usual setup in type-II 2HDM, we assign a $Z_2$ charge for each field to forbid tree level flavor-changing neutral currents and to simplify the scalar potential.

More explicitly, the quantum numbers of SM fermions and the two Higgs doublets $\Phi_1$ and $\Phi_2$ are listed in Table~\ref{table:2hdm}, and those of the fourth-generation fermions and the new scalar $\phi$ in Table~\ref{table:new}.  Gauge anomaly cancellation explicitly requires the relationship $9x_q+3x_{q4}=x_{\ell4}$.  In our numerical studies, we take $x_q=x_{q4}=1/3$ to be the same as their baryon number. Furthermore, the charge assignment of the SM singlet scalar field under $U(1)_X$ plays an important role to forbid the scalar $\phi$ to decay into SM particles and Higgs bosons, which makes $\phi$ a suitable DM candidate. The DM phenomenology will be discussed later. Also, we assume the $U(1)_X$ to be spontaneously broken at a certain energy scale higher than the electroweak scale.  As a result, we have a new heavy gauge boson $Z'$ that is crucial in the scattering cross section between DM and nucleon.

Unlike the sequential fourth generation, the SM gauge charges of the left- and right-handed fourth-generation leptons are interchanged.  Such a setup can lead to an enhancement in the muon $g-2$ at the one-loop level due to the interactions between leptons and the signet scale field $\phi$, to be discussed in detail in the next section.

\begin{table}[t!]
\centering
\begin{tabular}{|l|l|l|l|l|l|l|l|l}
\hline\hline
\multicolumn{1}{|c|}{} & \multicolumn{6}{c|}{\textbf{SM fermions}}                                                                                                                                                                       & \multicolumn{2}{c|}{\textbf{Higgs}}                       
\\ \hline
\multicolumn{1}{|c|}{} & \multicolumn{1}{c|}{$Q_{L}$} & \multicolumn{1}{c|}{$u_{R}$} & \multicolumn{1}{c|}{$d_{R}$} & \multicolumn{1}{c|}{$L_{L}$} & $e_{R}$     & $\nu_{R}$   & $\Phi_1$     & \multicolumn{1}{l|}{$\Phi_2$}     
\\ \hline
$SU(3)_C$              & $\textbf{3}$                                           & $\textbf{3}$                  & $\textbf{3}$                  & $\textbf{1}$                                             & $\textbf{1}$ & $\textbf{1}$ & $\textbf{1}$ & \multicolumn{1}{l|}{$\textbf{1}$} \\ \hline
$SU(2)_L$              & $\textbf{2}$                                           & $\textbf{1}$                  & $\textbf{1}$                  & $\textbf{2}$                                             & $\textbf{1}$ & $\textbf{1}$ & $\textbf{2}$ & \multicolumn{1}{l|}{$\textbf{2}$} \\ \hline
$U(1)_Y$               & 1/6                                                    & 2/3                           & $-1/3$                          & $-1/2$                                                     & $-1$           & 0            & 1/2          & \multicolumn{1}{l|}{1/2}          \\ \hline
$U(1)_X$               & $x_q$     & $x_q$      & $x_q$     & 0                                                        & 0            & 0            & 0            & \multicolumn{1}{l|}{0}            \\ \hline
$Z_2$                  & +                                                      & +                             & $-$                             & +                                                        & $-$            & +            & +            & \multicolumn{1}{l|}{$-$}            \\ 
\hline\hline
\end{tabular}
\caption{Quantum numbers of SM fermions in one generation and two Higgs doublets. 
}
\label{table:2hdm}
\end{table}
\begin{table}[t!]
\centering
\begin{tabular}{|l|l|l|l|l|l|l|l|}
\hline\hline
         & \multicolumn{6}{c|}{\textbf{Fourth generation fermions}}                  & \textbf{Scalar}      
\\\hline
         & $Q_{L4}$       & $u_{R4}$       & $d_{R4}$       & $L_{R4}$       & $e_{L4}$       & $\nu_{L4}$     & $\phi$          
\\\hline
$SU(3)_C$ & $\textbf{3}$ & $\textbf{3}$ & $\textbf{3}$ & $\textbf{1}$ & $\textbf{1}$ & $\textbf{1}$ & $\textbf{1}$ \\\hline
$SU(2)_L$ & $\textbf{2}$ & $\textbf{1}$ & $\textbf{1}$ & $\textbf{2}$ & $\textbf{1}$ & $\textbf{1}$ & $\textbf{1}$ \\\hline
$U(1)_Y$  & $1/6$        & $2/3$        & $-1/3$       & $-1/2$       & $-1$         & $0$          & $0$        \\\hline
$U(1)_X$  & $x_{q4}$ & $x_{q4}$  & $x_{q4}$  & $x_{\ell4}$          & $x_{\ell4}$          & $x_{\ell4}$          & $x_\phi$          \\\hline
$Z_2$  & $+$        & $+$        & $-$        & $+$          & $-$          & $+$          & $+$           \\\hline\hline
\end{tabular}
\caption{Quantum numbers of fourth-generation fermions and new scalar.}
\label{table:new}
\end{table}

With the assignment $x_\phi=x_{\ell4}$, the interaction Lagrangian for the SM leptons, the four-generation leptons and the scalar boson $\phi$ can be written down as
\begin{align}
\begin{split}
\mathcal{L}_{\text{Y}} 
=& 
-\left[f^{L}_{4i} \bar{L}_{R4}L_{Li} +f^{e}_{4i} \bar{e}_{L4} e_{Ri} + f^{\nu}_{4i} \bar{\nu}_{L4} \nu_{Ri} \right]\phi  
\\
& + {\rm h.c.}
~,
\end{split}
\label{eq:yukawa}
\end{align}
where the flavor index $i = e, \mu, \tau$.  We denote the mass of $e_4$ by $m_{e_4}$.  Because of the $U(1)_X$ charges, such interactions are always lepton flavor-violating.  For simplicity and concreteness, we will take the Yukawa couplings
\begin{align}
f^L_{4i}=f^e_{4i}=f^\nu_{4i} \equiv f_{4i}
\label{eq:yukawa-coupling-assumption}
\end{align}
to be real in the following numerical studies.
As far as the muon anomalous magnetic moment is concerned, only $f^L_{4\mu}$ and $f^e_{4\mu}$ are relevant.

Following the notations in Ref.~\cite{Branco:2011iw}, the two Higgs doublets $\Phi_1$ and $\Phi_2$ are parameterized as
\begin{equation}
\Phi_i = \begin{pmatrix}
\phi^+_i\\ 
(v_i + \rho_i + i\eta_i)/\sqrt{2}
\end{pmatrix}
~,
\end{equation}
where $ i=1,2$, $v_1 = v \cos \beta$ and $v_2 = v\sin\beta$ with $v = 246$~GeV. 
The physical neutral scalars $h$ and $H$ can be expressed as linear combinations of $\rho_1$ and $\rho_2$:
\begin{equation}
\begin{aligned}
h &= \rho_1 \sin \alpha - \rho_2 \cos\alpha
~, \\
H &= -\rho_1 \cos\alpha - \rho_2 \sin\alpha
~,
\end{aligned}
\end{equation}
where $\alpha$ is a mixing angle among the neutral components.
The scalar potential in the model includes two parts:
\begin{widetext}
\begin{equation}
\begin{aligned}
V_{\text{2HDM}} 
&= m^2_{11} \abs{\Phi_1}^2 + m^2_{22} \abs{\Phi_2}^2 - m^2_{12}(\Phi^\dagger_1 \Phi_2 + \Phi_1 \Phi^\dagger_2) 
\\
& \qquad
+ \frac{\lambda_1}{2} \abs{\Phi_1}^4 + \frac{\lambda_2}{2} \abs{\Phi_2}^4 + \lambda_3 \abs{\Phi_1}^2 \abs{\Phi_2}^2 
+ \lambda_4 \abs{\Phi^\dagger_1 \Phi_2}^2 + \frac{\lambda_5}{2} [(\Phi^\dagger_1 \Phi_2)^2 + (\Phi_1 \Phi^\dagger_2)^2]
~, \\
V_\phi & = m^2_0 \abs{\phi}^2 + \lambda_\phi \abs{\phi}^4 + \kappa_1 \abs{\phi}^2 \abs{\Phi_1}^2 +\kappa_2\abs{\phi}^2 \abs{\Phi_2}^2 
~,
\end{aligned}
\end{equation}
\end{widetext}
where $V_{\text{2HDM}}$ is the scalar potential for the usual 2HDMs with a $Z_2$ symmetry softly broken by the $m_{12}^2$ terms and $V_\phi$ contains terms involving purely $\phi$ as well as mixing with $\Phi_{1,2}$.  Here we assume that the new scalar field $\phi$ does not develop a vacuum expectation value (VEV).
The singlet scalar mass $m_\phi$ and the coupling strength of the $h$-$\phi$-$\phi$ and $H$-$\phi$-$\phi$ vertices, denoted respectively by $\lambda_h$ and $\lambda_H$, can be derived from the above scalar potential to be
\begin{align}
\begin{split}
m^2_\phi &= m^2_0 + \frac{1}{2}\kappa_1 v^2 \cos^2 \beta + \frac{1}{2}\kappa_2 v^2 \sin^2\beta
~, \\
\lambda_h v &= (-\kappa_1 \sin\alpha \cos \beta + \kappa_2 \cos\alpha \sin \beta)v
~, \\
\lambda_H v&= (\kappa_1 \cos\alpha \cos\beta + \kappa_2 \sin\alpha \sin\beta)v
~.
\end{split}
\end{align}
We will consider the scenario where $\lambda_H \ll 1$ and, as a consequence, the coupling 
\begin{equation}
\lambda_h \simeq  \kappa_2 \frac{\sin\beta}{\cos\alpha}
~. 
\end{equation}

The Yukawa couplings of up-type fermions $y^h_u$ and down-type fermions $y^h_d$ are
\begin{align}
\begin{split}
y^h_u &= \frac{\cos\alpha}{\sin \beta} = \cos(\beta - \alpha) \cot \beta + \sin (\beta - \alpha)
~, \\
y^h_d &= -\frac{\sin\alpha}{\cos\beta} = \sin (\beta-\alpha) -\cos(\beta-\alpha)\tan \beta
~.
\end{split}
\end{align}
Note that the wrong-sign scenario can be realized when $\cos(\beta-\alpha)\lesssim 1/\tan\beta$ and $\tan\beta \agt 2$~\cite{Das:2017mnu,Han:2017etg}.  Unless otherwise specified, we adopt the following benchmark mass spectrum similar to the one given in Ref.~\cite{Das:2017mnu} for the numerical results presented in the following sections: 
\begin{eqnarray*}
m_{u_4}&=&550~{\rm GeV} ~,~~
m_{d_4}=510~{\rm GeV} ~,~~
m_{e_4}=400~{\rm GeV} ~,
\\
m_{\nu_4}&=&400~{\rm GeV} ~,~~
m_H= 400~{\rm GeV} ~,~~
m_A= 810~{\rm GeV} ~,
\\
m_{H^+}&=&600~{\rm GeV} ~.
\end{eqnarray*}
%

\section{Muonic observables and constraints}
\label{sec:gm2}

\begin{figure}[t!h]
\centering
\includegraphics[scale=0.45]{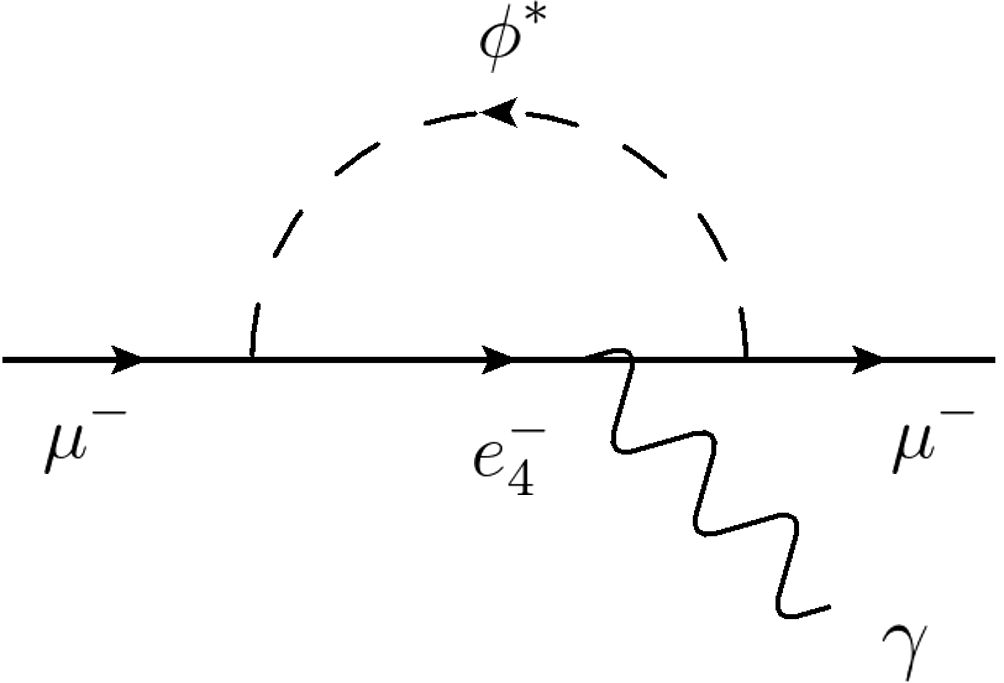}
\caption{One-loop contribution to muon $g-2$ anomaly in the model.}
\label{fig:feyn}
\end{figure}

In additional to the SM calculations, the muon anomalous magnetic dipole moment $a_\mu$ receives a major contribution from the one-loop diagram mediated by the fourth-generation charged lepton $e_4$ and the scalar boson $\phi$ under our setup, as shown in Fig.~\ref{fig:feyn}, 
Explicitly, we have~\cite{Jegerlehner:2009ry}  
\begin{equation}
\Delta a_\mu 
= 
\frac{\abs{f_{4 \mu}}^2}{8\pi^2} \int^1_0 dx \frac{x^2 [ m_\mu^2(1-x)+ m_\mu m_{e_4} ] }{(1-x)(m_\phi^2- m_\mu^2x) + m_{e_4}^2x}
~.
\end{equation}
The term involving $m_{e_4}$ gives the major contribution.  Other diagrams, including those mediated by CP-even, CP-odd and charged Higgs bosons, give a result about four orders of magnitude smaller and are neglected.  It is easy to see that $\Delta a_\mu$ depends upon three parameters of the model: $f_{4\mu}$, $m_{e_4}$ and $m_\phi$.

\begin{figure}[t!h]
\centering
\subfigure[]{
    \includegraphics[scale=0.4]{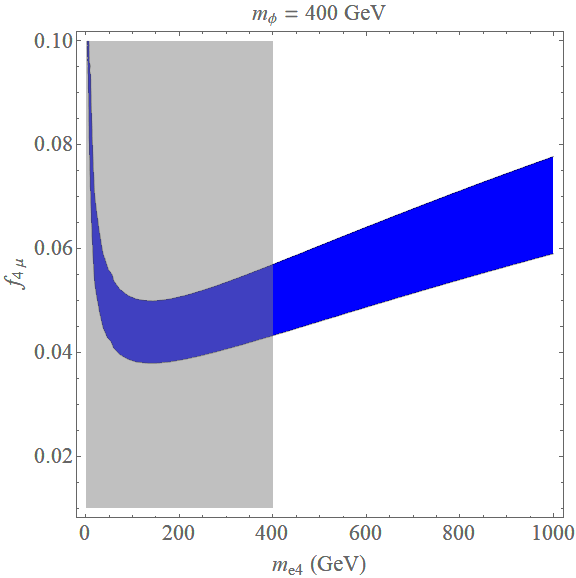}
    \label{fig:gm2a}
}
\smallskip
\subfigure[]{
    \includegraphics[scale=0.4]{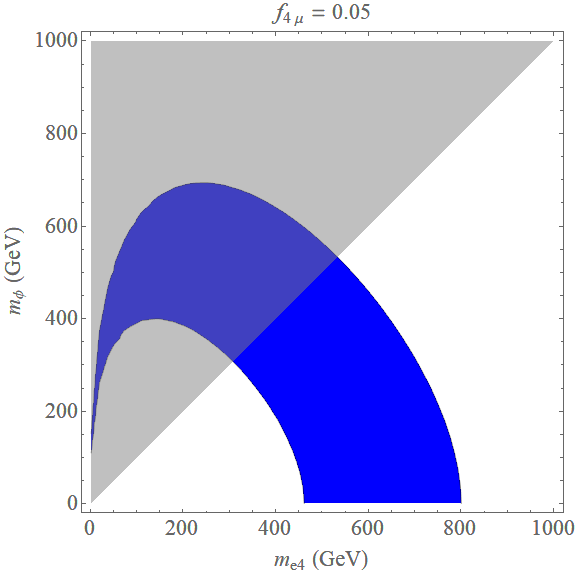}
    \label{fig:gm2b}
}
\caption{Parameter space favored by $\Delta a_\mu$ (a) in the $m_{e_4}$-$f_{4\mu}$ plane with $m_{\phi} = 400$~GeV and (b) in the $m_{e_4}$-$m_\phi$ plane with $f_{4\mu} = 0.05$.  The grey region in both plots is excluded by the stability requirement of the DM candidate $\phi$.}
\label{fig:gm2fit}
\end{figure}

In Fig.~\ref{fig:gm2fit}, we show the parameter space in blue band that fits the observed $\Delta a_\mu$ at $1\sigma$ level. The grey region is excluded by the requirement of $\phi$ being the dark matter candidate and its decay to $e_4$ being kinematically forbidden. 
In Fig.~\ref{fig:gm2a}, we fix the mass of $\phi$ to be $400$~GeV, and the coupling strength $f_{4\mu}$ is then found to fall between $0.04$ and $0.08$ for the heavy charged lepton up to about $1$ TeV. 
With $f_{4\mu}=0.05$ fixed in Fig.~\ref{fig:gm2b}, we observe that the mass of $e_4$ is preferred to be in the window of $300$~GeV to $ 800$~GeV.  As shown in the next section, the DM mass $m_\phi$ will be further restricted by the relic density measurement and direct detection bound.

In addition to offering a possible account of the muon $g-2$ anomaly, the interactions in Eq.~(\ref{eq:yukawa}) with the assumption in Eq.~\eqref{eq:yukawa-coupling-assumption} also lead to lepton flavor-violating processes that should be controlled to be compatible with observed limits.  Here we consider the following flavor-changing rare decays of muon and tau: $\mu^- \rightarrow e^-\gamma$, $\tau^- \rightarrow e^-\gamma$, $\tau^- \rightarrow \mu^-\gamma$ and $\mu^- \rightarrow e^-e^-e^+$.
The first three processes occur through the one-loop diagrams analogous to Fig.~\ref{fig:feyn} with the initial- and final-state muon replaced by the appropriate leptons, while the last one involves both penguin and box diagrams.

As an explicit example, the branching ratio of $\mu^- \rightarrow e^-\gamma$ is approximately given by~\cite{Calibbi:2017uvl}
\begin{align}
\begin{split}
\label{eq:br}
&
\text{BR}(\mu^- \rightarrow e^- \gamma) 
\simeq \frac{\Gamma(\mu\to e\gamma)}{\Gamma(\mu\to e \nu \bar{\nu})}
\\
&= 
\frac{m_\mu^3}{16\pi} \left( \abs{\sigma_L}^2 + \abs{\sigma_R}^2 \right) \frac{192\pi^3}{G^2_F m^5_\mu}
~,
\end{split}
\end{align}
where the Fermi constant $G_F = 1.166 \times 10^{-5}~ \text{GeV}^{-2}$ and 
\begin{align}
\begin{split}
&
\sigma_L = \sigma_R \\
&\simeq 
i(4\pi)^2 e f^*_{4e} f_{4\mu} \int^1_0  dz \int^{1-z}_0 dy \frac{m_{e_4}(z-1)}{m^2_\phi z + m_{e_4}^2(1-z)}
~.
\end{split}
\label{eq:sigma}
\end{align}
The branching ratios of $\tau\to \mu \gamma$ ($\tau\to e\gamma$) can be obtained by replacing $m_\mu$ in Eq.~(\ref{eq:br}) with $m_\tau$ and replacing $f_{4e}$ ($f_{4\mu}$) with $f_{4\tau}$ in Eq.~(\ref{eq:sigma}).  The analytical result of $\text{BR}(\mu\to e e e)$ is more involved and given in the Appendix.

Currently, the upper bounds on the charged lepton flavor-violating decay modes are~\cite{TheMEG:2016wtm,Patrignani:2016xqp,Pruna:2018egr}:
\begin{equation}
\begin{aligned}
\text{BR}(\mu^- \rightarrow e^-\gamma) &< 4.2 \times 10^{-13}
~, \\
\text{BR}(\tau^- \rightarrow e^-\gamma) &< 3.3 \times 10^{-8}
~, \\
\text{BR}(\tau^- \rightarrow \mu^-\gamma) &< 4.4 \times 10^{-8}
~, \\
\text{BR}(\mu^- \rightarrow e^-e^-e^+) &< 1\times 10^{-12}
~,
\end{aligned}
\label{eq:cfg}
\end{equation}
all quoted at the $90\%$ confidence level.  With $f_{4i}$ assumed to be real, the bounds in Eq.~(\ref{eq:cfg}) impose stringent constraints on the coupling products $f_{4e}f_{4\mu}$, $f_{4e}f_{4\tau}$ and $f_{4\tau}f_{4\mu}$.  To satisfy the current limits, $f_{4e}$ and $f_{4\tau}$ are found respectively to be five and three orders of magnitude smaller than $f_{4\mu}\simeq 0.05$.

Because of the above-mentioned hierarchy among the $f_{4i}$ couplings, the box diagram for the $\mu\to eee$ decay is negligible as compared to the penguin diagram because it is proportional to $f_{4\mu}(f_{4e})^3$ while the latter involves just $f_{4\mu}f_{4e}$. 
Furthermore, the branching ratio of $\mu^- \rightarrow e^-e^-e^+$ has the quasi model-independent relation with $\mu^- \rightarrow e^-\gamma$~\cite{Kuno:1999jp,Feldmann:2016hvo}:
\begin{equation}
\text{BR}(\mu^- \rightarrow e^-e^-e^+)/\text{BR}(\mu^- \rightarrow e^-\gamma) \approx 0.006
~.
\end{equation} 
Numerically, we have found the constraints from the radiative decays stronger than that from the three-body decay.  
The couplings $f_{4e}$ and $f_{4\tau}$ are found to be less contained in other charged lepton flavor-changing decays, $\tau \to 3\mu$, $\tau \to 3e$ and $\tau \to \mu ee$.

\begin{figure}[t!h]
\centering
\includegraphics[scale=0.5]{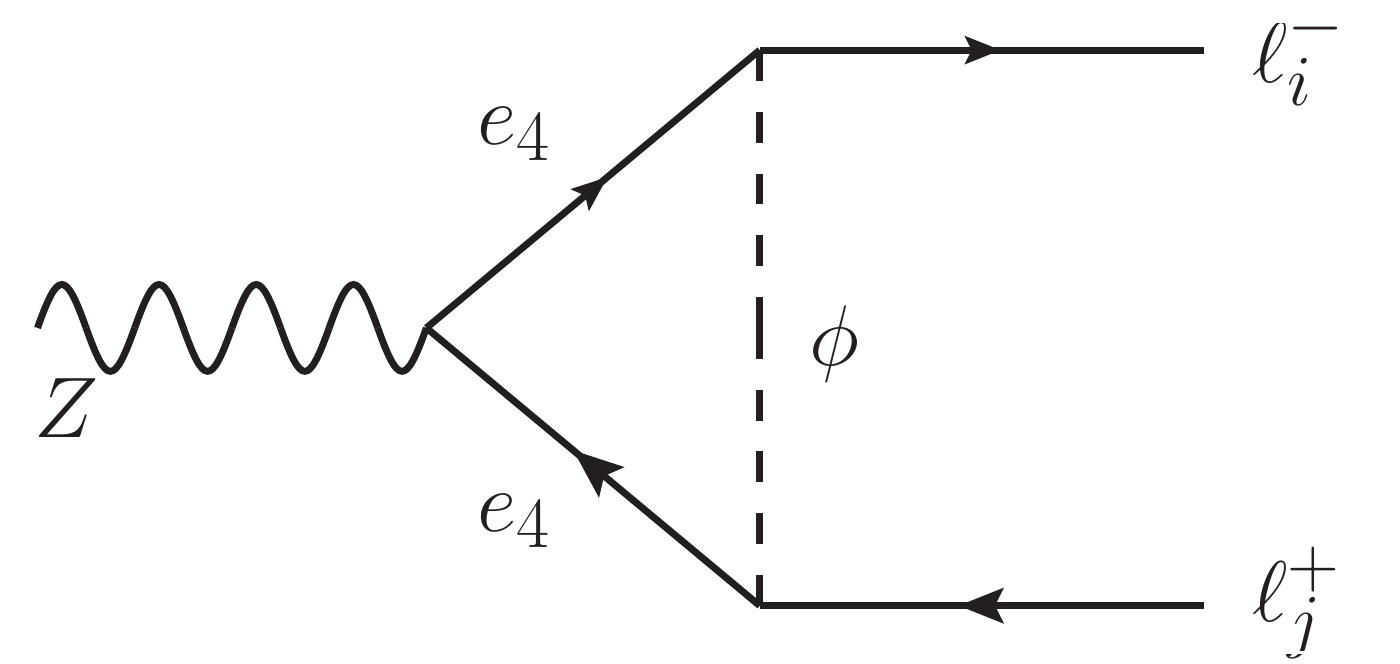}
\caption{A representative Feynman diagram of charged lepton flavor-violating decays of the $Z$ boson in the model.}
\label{fig:zee}
\end{figure}

Besides, lepton flavor-violating decays of the $Z$ boson can be induced at the one-loop level due to the interactions in Eq.~(\ref{eq:yukawa}), as illustrated in Fig.~\ref{fig:zee}.  The branching ratio of $Z \rightarrow \ell_i^{\pm} \ell_j^{\mp}$ is given by~\cite{Illana:2000ic}
\begin{align}
\begin{split}
\label{k}
&
\text{BR}(Z \rightarrow \ell_i^{\pm} \ell_j^{\mp}) 
\\
&= \frac{m_Z}{12\pi \Gamma_Z} \left[ 2 \left( f_V^2 + f_A^2 \right) + \left( f_M^2 + f_E^2 \right) m^2_Z \right]
~,
\end{split}
\end{align}
where $\ell_{i,j}=e,\mu,\tau$ and the $Z$ boson total decay width $\Gamma_Z \simeq g^2/(4\pi \cos^2\theta_W) m_Z$, $g$ is the $SU(2)_L$ gauge coupling constant, and $\theta_W$ is the weak mixing angle.  The couplings in Eq.~\eqref{k} are given by
\begin{align}
\begin{split}
f_V =& \left[ g' \left( -\frac{1}{4} + \sin^2\theta_W \right) \right]
\frac{f_{4j} f^*_{4i}}{(4\pi)^2} \left[ I_1 + I_2 \right]
~,
\\
f_A =& \frac{g'}{4} \frac{f_{4j} f^*_{4i}}{(4\pi)^2} \left[ I_1 + I_2 \right]
~,
\\
f_M =& \left[ g' \left( -\frac{1}{4} + \sin^2\theta_W \right) \right] 
\frac{f_{4j} f^*_{4i}}{(4\pi)^2} I_1
~,
\\
f_E =& \frac{g'}{4} \frac{f_{4j} f^*_{4i}}{(4\pi)^2} I_1
~,
\end{split}
\end{align}
where $g' \equiv g / (\cos\theta_W)$ and
\begin{widetext}
\begin{align}
\begin{split}
I_1
&\equiv 
\int^1_0 dz \int^{1-z}_0 dy \frac{m_{4}^2 + y(1-y-z)m^2_Z}{-y(1-y-z) m_Z^2 +(1-z)m_{4}^2 + zm_\phi^2}
~,
\\
I_2
&\equiv 
\int^1_0 dz z \ln\left[ z m_{4}^2 + (1-z)m^2_\phi \right]
- \int^1_0 dz \int^{1-z}_0dy \ln\left[ (1-z)m_{4}^2 + zm^2_\phi -y(1-y-z)m^2_Z \right] 
~.
\end{split}
\end{align}
\end{widetext}
Currently, the experimental upper bounds~\cite{Illana:2000ic}
\begin{equation}
\begin{aligned}
& \text{BR}(Z \rightarrow e^{\pm} \mu^{\mp}) < 1.7 \times 10^{-6}\\
& \text{BR}(Z \rightarrow e^{\pm} \tau^{\mp}) < 9.8 \times 10^{-6}\\
& \text{BR}(Z \rightarrow \mu^{\pm} \tau^{\mp}) < 1.2 \times 10^{-5}
\end{aligned}
\label{eq:zeelimit}
\end{equation}
constrain $f_{4e}f_{4\mu}$, $f_{4e}f_{4\tau}$ and $f_{4\mu}f_{4\tau}$, respectively. 
To explain the muon $g-2$ anomaly, the preferred value of $f_{4\mu}$ is around $0.05$ when masses of $\phi$ and $e_4$ are of ${\cal O}(100)$~GeV as seen in Fig.~\ref{fig:gm2fit}.  Given this parameter space preferred by the observed $\Delta a_\mu$, the upper bounds on $f_{4e}$ and $f_{4\tau}$ extracted from Eq.~(\ref{eq:zeelimit}) turn out to be much less stringent than the bounds obtained from the rare charged lepton decays in Eq.~(\ref{eq:cfg}).

\section{Dark Matter}
\label{sec:dm}

In this section, we study the phenomenology of DM candidate $\phi$, including its relic density and constraint from the null result of direct search.
\begin{figure}[t!]
\begin{center}
\includegraphics[scale=0.4]{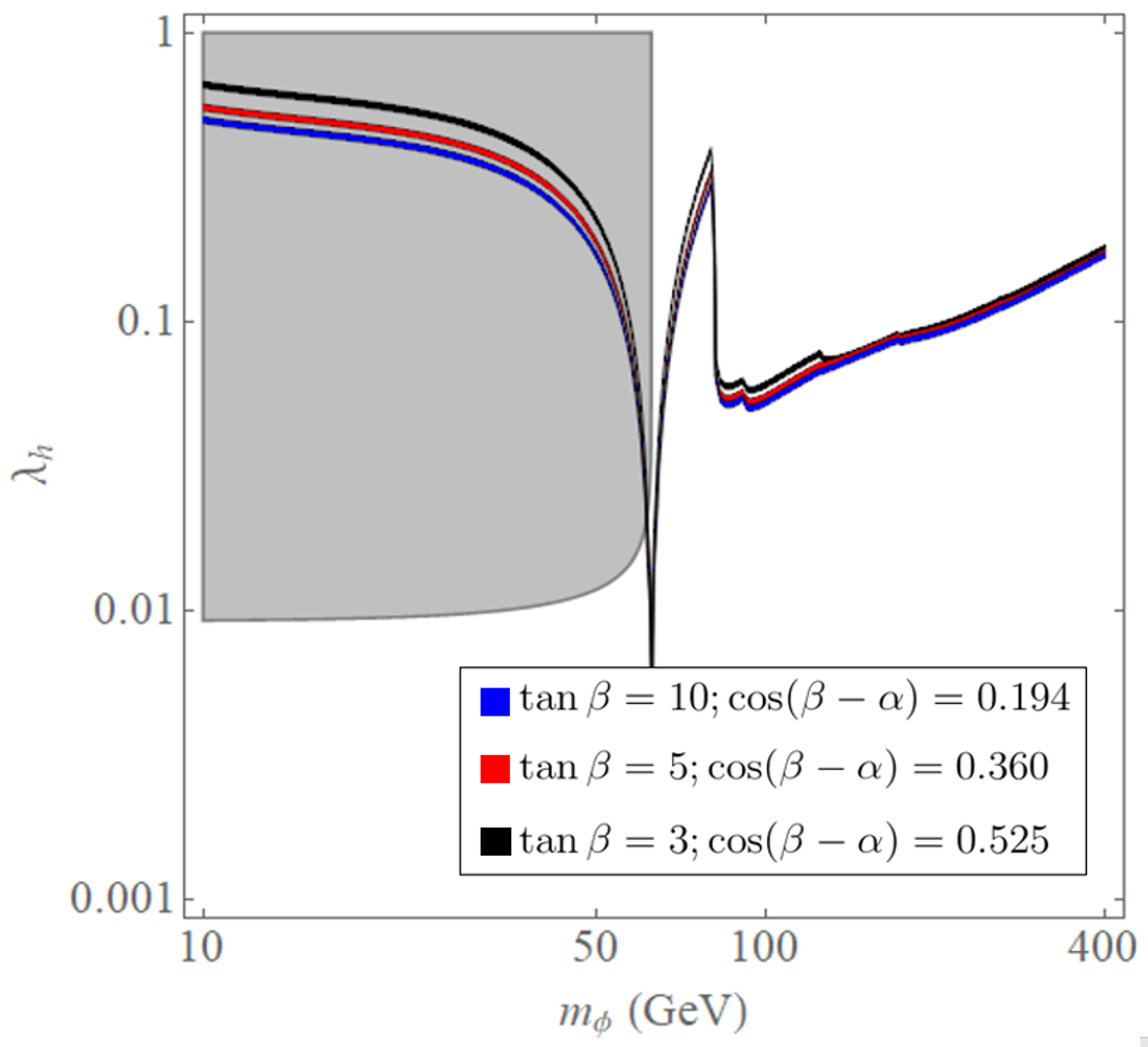}
\caption{Parameter space in the $m_\phi$-$\lambda_h$ plane that fits the observed current DM relic density $\Omega h^2 = 0.1200 \pm 0.0060$ at the $5\sigma$ confidence level.  The mass of $Z'$ is fixed at $1$~TeV.  The three colored bands have different values of $\tan\beta$ and $\cos(\beta - \alpha)$, as shown in the plot.  The grey region is excluded by the measured Higgs invisible decay width. }
\label{fig:relic}
\end{center}
\end{figure}

The present dark matter relic density is determined by its cross section of annihilation to SM particles. Two $\phi's$ can annihilate into several possible final states of SM particles in pairs, including leptons, quarks, gauge bosons and Higgs bosons.  The first case involves a fourth-generation lepton as the mediator in the $t$-channel process.  However, this channel gives a tiny contribution when we take into account the rare decay limits of charged leptons discussed previously.  Therefore, the annihilation cross section is dominated by diagrams mediated by the 125-GeV Higgs boson $h$ and the $Z'$ boson, between which the former plays a dominant role.

Fig.~\ref{fig:relic} shows the parameter space in the $m_\phi$-$\lambda_h$ plane that renders the observed DM relic density $\Omega h^2= 0.1200 \pm 0.0012$~\cite{Aghanim:2018eyx}, where $\lambda_h$ parameterizes the coupling strength of the $h$-$\phi$-$\phi$ vertex which reads as $2 \lambda_h m_W/g$. The black, red and blue bands are respectively the results with $\tan\beta = 3$, $5$ and $10$, and the angle $\alpha$ is chosen to comply with the wrong-sign scenario. The colored bands show a few structures due to various kinematical reasons.  As the DM mass is close to half of the Higgs boson mass, the sharp drop reflects the resonance enhancement in the annihilation cross section. Similarly, there are then a few more drops as the DM mass crosses the thresholds to annihilate into a pair of $W$ bosons, $Z$ bosons, Higgs bosons and top quarks, respectively.  In this plot, we also draw a grey area that is excluded by the upper limit of Higgs boson invisible decay branching ratio $\text{BR}(h\to \phi\phi)\lesssim 0.24$ \cite{Khachatryan:2016whc}.

We now turn to the spin-independent elastic scattering between the DM and a nucleon, related to direct search experiments.  
The scattering process receives contributions that features the exchanges of Higgs and $Z'$ bosons.
In the low-energy approximation, the DM-nucleon (proton or neutron) cross section can be expressed as
\begin{equation}
\sigma^{\rm SI}_{\phi p(n)} = \sigma^{{\rm SI}-h}_{\phi p(n)} + \sigma^{{\rm SI}-Z^\prime}_{\phi p(n)}
~,
\end{equation}
with negligible interference between the two diagrams due to the large mass separation between $h$ and $Z'$.
The contribution from the Higgs-mediated diagram reads 
\begin{equation}
\sigma^{{\rm SI}-h}_{\phi p(n)} = \frac{m^2_{p(n)}}{4\pi \left( m_\phi + m_{p(n)} \right)^2} 
\left( f^{p(n)}_S \right)^2
~,
\end{equation}
where $m_p$ and $m_n$ are the masses of proton and neutron, respectively, and
\begin{equation}
\frac{f_S^{p(n)}}{m_p} = \sum_{q=u,d,s} \frac{\alpha^{S}_q}{m_q} f^{p(n)}_{Tq} + \frac{2}{27} f^{p(n)}_{TG} \sum_{q=c,b,t} \frac{\alpha^{S}_q}{m_q}
~.
\end{equation}
In the above equation, $f^{p(n)}_{Tq}$ represent the contributions of light quarks to the mass of the proton (neutron) and $\alpha^S_q \approx -i \lambda_h m_q/ m^2_h$ is the effective coupling after integrating out the Higgs boson in the low-energy approximation.  The second term represents the interaction of DM with the gluon scalar density in the nucleon, with $f^{p(n)}_{TG} = 1 - \sum_{q=u,d,s} f^{p(n)}_{Tq}$.  Values of these form factors used in the numerical analysis are taken as~\cite{Junnarkar:2013ac,Hoferichter:2015dsa,Alarcon:2011zs,Alarcon:2012nr,Cheng:2012qr}
\begin{align}
\begin{split}
f^p_{Tu} & = 0.0208 \pm 0.0015 ~,~~ f^n_{Tu} = 0.0189 \pm 0.0014 ~,\\
f^p_{Td} & = 0.0411 \pm 0.0028 ~,~~ f^n_{Td} = 0.0451 \pm 0.0027 ~,\\
f^p_{Ts} & = 0.043 \pm 0.011 ~,~~ f^n_{Ts} =0.043 \pm 0.011 ~.
\end{split}
\end{align}

The other contribution to the spin-independent DM-nucleon scattering comes from the $Z^\prime$-mediated diagram.  This cross section is given by
\begin{equation}
\sigma^{{\rm SI}-Z^\prime}_{\phi p(n)} 
= \frac{(3x_q)^2 x_\phi^2 e^{\prime 4} m^2_\phi m^2_{p(n)}}
{\pi \left( m_\phi + m_{p(n)} \right)^2 m^4_{Z^\prime}},
\end{equation}
where $e'$ denotes the gauge coupling strength of the $U(1)_X$ group, and $x_q$ and $x_\phi$ are the $U(1)_X$ charges of SM quarks and dark matter, respectively.

\begin{figure}[t!]
\centering
 \includegraphics[scale=0.4]{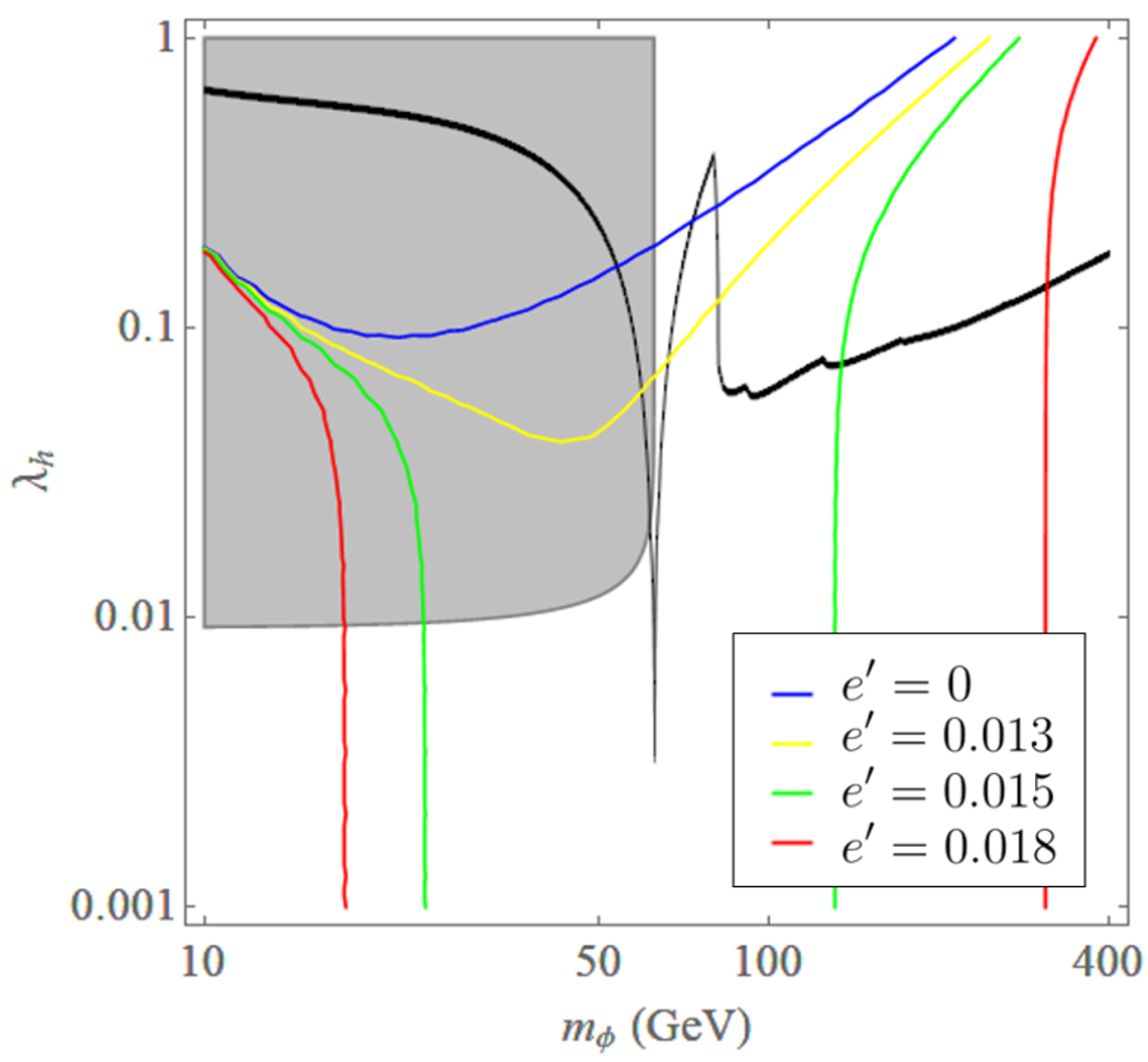}
\caption{Regions above the colored curves are excluded by the XENON1T measurement~\cite{Aprile:2017iyp}.  Different colors are for different values of the gauge coupling $e^\prime$.  The black band satisfies the current DM relic density with $\tan\beta = 3$ and $\cos(\beta - \alpha) = 0.525$.  The grey region is excluded by the measured Higgs invisible decay width.}
\label{fig:es_tan}
\end{figure}

The blue curve in Fig.~\ref{fig:es_tan} shows the upper limit of elastic scattering cross section that is consistent  with the null result of XENON1T experiment~\cite{Aprile:2017iyp} if only the Higgs-mediated diagram is considered.  The black band and the grey region are quoted from the results with correct dark matter relic density with $\tan\beta = 3$ and $\cos(\beta - \alpha) = 0.525$ and exclusion by Higgs invisible decay as given in Fig.~\ref{fig:relic}.  Most of the region with $m_\phi$ heavier than $m_h/2$ are allowed, except for a small mass window around $m_\phi\simeq 80$~GeV.

Other curves in yellow, green and red represent the constraints for different $U(1)_X$ gauge coupling strength $e'=0.013,~0.015$ and $0.018$, respectively.  Similarly, the region above each curve is excluded.  Clearly, the elastic cross section is very sensitive to the $U(1)_X$ gauge coupling strength, and the $Z^\prime$-mediated diagram can easily dominate over the Higgs boson contribution when the new gauge coupling is turned on.  For example, $m_\phi \lesssim 300$~GeV is excluded in the case of $e'=0.018$.

\begin{figure}[t!]
\centering
 \includegraphics[scale=0.4]{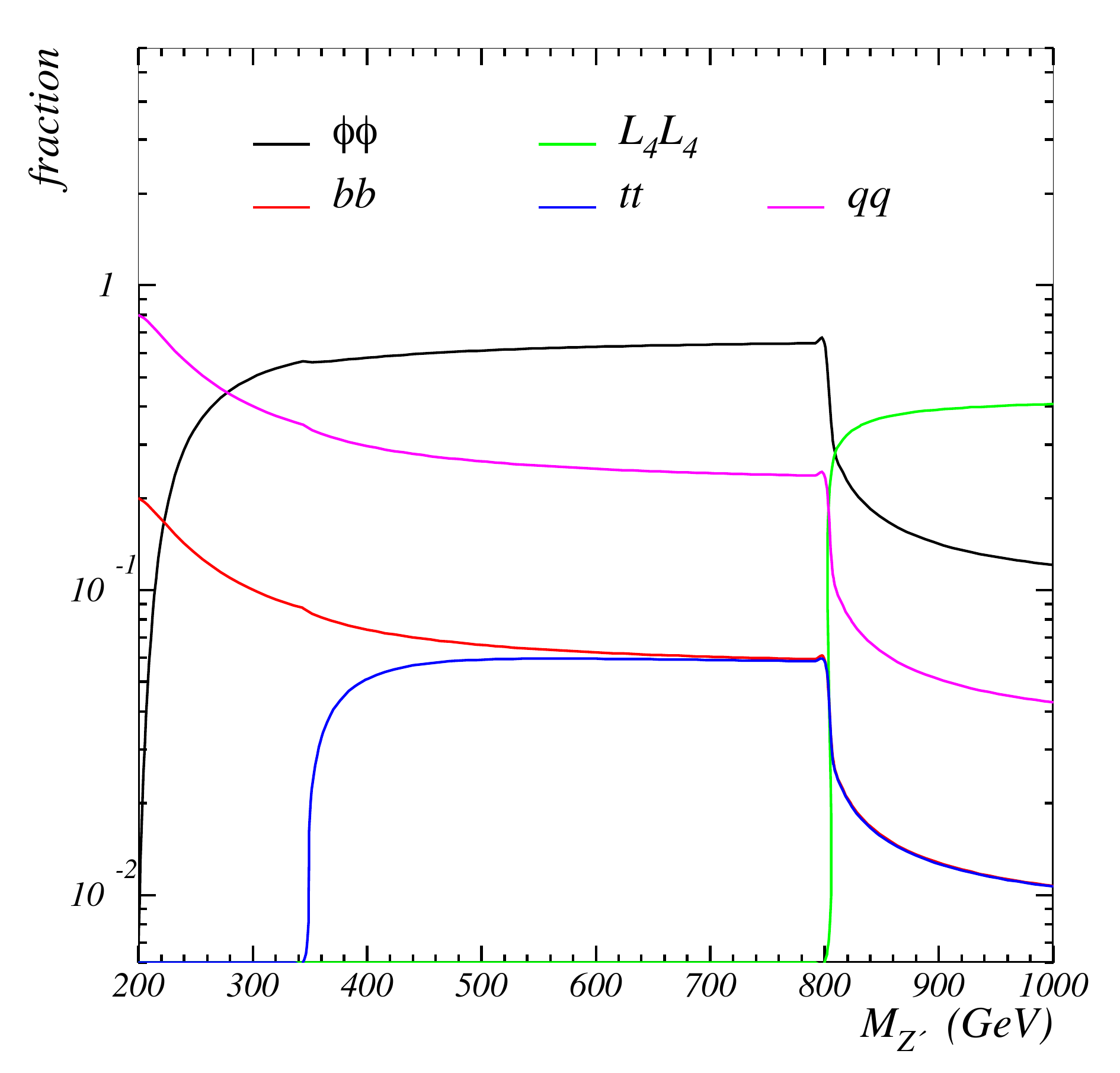}
\caption{Decay branching ratios of $Z'$, where  $L_4=e_4$ or $\nu_4$, and $Z'\to qq$ is the sum of the decay branching ratios to the first two generations of SM quarks. Here, we take masses of fourth-generation leptons and dark matter to be $m_{\ell_4}=400$~GeV and $m_\phi=100$~GeV, respectively.}
\label{fig:zpdecay}
\end{figure}

Finally, we make a brief comment on the phenomenology of $Z'$ in the model at hadron colliders.  Since the $Z'$ boson couples to quarks, it can be searched through the dijet resonance. However, the usual resonance search in dilepton channel would not apply since the SM leptons are $U(1)_X$ charge neutral. And we have checked that the numerical results presented in this work satisfy the $Z^\prime$ lower bound set by the LHC dijet resonance search~\cite{Lee:2011jk}. 
Furthermore, since $Z'$ couples to the fourth-generation leptons and dark matter as well,  it mainly decays into dark matter particles if $m_{Z'}\lesssim 2 m_{\ell_4}$. When the decay channels to fourth-generation leptons are kinematically allowed, the branching ratios to $\ell_4$ become dominant, as we can see in Fig.~\ref{fig:zpdecay}.
Therefore, the dijet branching ratio will be suppressed. The signatures at the LHC would be mono-jet with large missing energy for a light $Z'$. For a heavy $Z'$ that decays dominantly into two fourth-generation charged leptons, one should focus on the signature of two charged leptons with missing energy.

\section{Conclusions \label{sec:conclude}}

The measurement of muon $g-2$, with a $\sim 3\sigma$ deviation away from the present SM predictions, may present hints of new physics. Furthermore, evidence showing the existence of dark matter certainly call for an extension of SM. In this paper, we have considered a type-II two-Higgs doublet model with an additional generation of fermions, a SM singlet scalar and a new $U(1)_X$ gauge group.         
With non-canonical $SU(2)_L$ charge assignments to the fourth-generation leptons, we showed that the additional contributions of fourth-generation charged leptons with the singlet scalar in the loop could fix the tension between muon $g-2$ measurement and theory predictions. In addition, flavor-changing rare decays of charged leptons and $Z$ boson could be induced in this model. And we have checked that the parameter space to explain muon $g-2$ anomaly is consistent with these constraints.

With the assumed $U(1)_X$ charges of fields in this model, the singlet scalar was found to be a suitable dark matter candidate that fits the observed relic density today. 
However, the LHC data of Higgs invisible decay excludes the possibility of dark matter being lighter than $m_h/2$. Furthermore, the scattering cross section between dark matter and nucleon is sensitive to the gauge coupling strength of $U(1)_X$. As a result, the null results in the direct search of dark matter experiments impose stringent constraints on dark matter mass even when the $U(1)_X$ gauge coupling is small.

\section*{Acknowledgments}
C.~W.~C. would like to thank the hospitality of the New High Energy Theory Center at Rutgers University where part of this work was done.
This research was supported in part by the Ministry of Science and Technology of Taiwan under Grant No.\ MOST 105-2112-M-003-010-MY3 (C.~R.~C.) and MOST 104-2628-M-002-014-MY4 (C.~W.~C.).

\appendix
\section{Details of $\mu\to3e$}

There are two kinds of Feynman diagrams that contribute to $\mu\to 3e$ decay at one-loop level: penguin diagram and box diagram, as shown in Fig.~7.
\begin{figure}[ht]
\begin{center}
\includegraphics[scale=0.22]{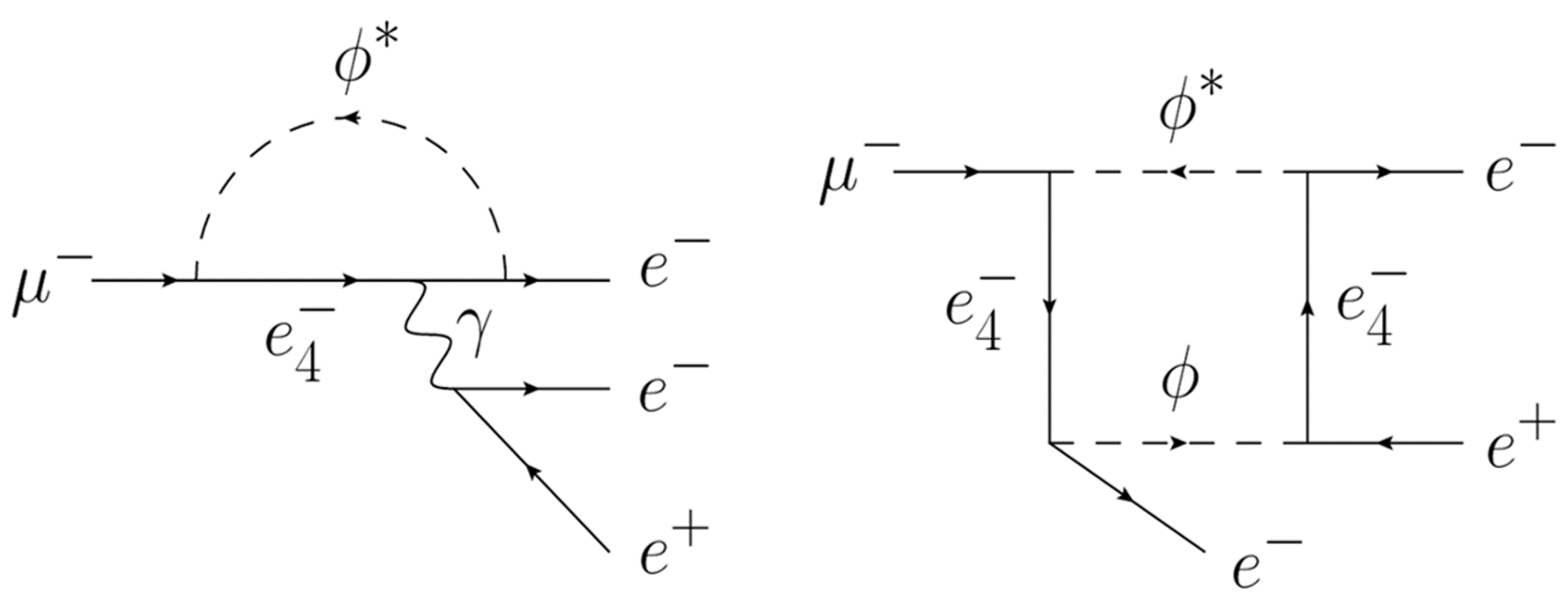}
\label{fig:muto3e}
\caption{Representative Feynman diagrams of $\mu^-\to e^+e^-e^-$: penguin diagram on the left and box diagram on the right.}
\end{center}
\end{figure}
Following the notations in Refs.~\cite{Kuno:1999jp,Feldmann:2016hvo}, the effective Lagrangian can be described as 
\begin{align}
&
\mathcal{L}_{\mu\to3e}
\\\nonumber
&=\mathcal{L}_{\mu\to e \gamma}\\\nonumber
&+ g_1 (\bar{\ell}_e P_R \ell_\mu)(\bar{\ell}_eP_R\ell_e) + g_2 (\bar{\ell}_e P_L \ell_\mu)(\bar{\ell}_eP_L \ell_e)\\\nonumber
&+ g_3 (\bar{\ell}_e \gamma^\nu P_R \ell_\mu)(\bar{\ell}_e\gamma_\nu P_R\ell_e) + g_4 (\bar{\ell}_e \gamma^\nu P_L \ell_\mu)(\bar{\ell}_e\gamma_\nu P_L\ell_e)\\\nonumber
&+g_5 (\bar{\ell}_e \gamma^\nu P_R \ell_\mu)(\bar{\ell}_e\gamma_\nu P_L\ell_e) + g_6 (\bar{\ell}_e \gamma^\nu P_L \ell_\mu)(\bar{\ell}_e\gamma_\nu P_R\ell_e)\\\nonumber
&+ {\rm h.c.}
~, 
\end{align}
where 
\begin{align}
\begin{split}
\mathcal{L}_{\mu\to e\gamma}
&= A_R m_\mu F_{\rho\delta}(\bar{\ell}_e \sigma^{\rho\delta}P_R\ell_\mu)
\\
&\qquad + A_L m_\mu F_{\rho\delta}(\bar{\ell}_e \sigma^{\rho\delta}P_L\ell_\mu)
~.
\end{split}
\end{align}
Note that, with this definition, the branching ratio of $\mu \to e\gamma$ reads as
\begin{align}
\begin{split}
\text{BR}(\mu\to e\gamma) 
&= \frac{\Gamma(\mu\to e\gamma)}{\Gamma(\mu\to e\nu\bar{\nu})}
\\
&= \frac{m_\mu^5}{4\pi \Gamma(\mu\to e\nu\bar{\nu})}(|A_R|^2+|A_L|^2)
~.
\end{split}
\end{align}
Therefore, the branching ratio of $\mu\to 3 e$ can be expressed as
\begin{align}
\begin{split}
&
{\rm BR(\mu\to3e)}
\\
&= \frac{m_\mu^5}{1536\pi^3 \Gamma(\mu\to e\nu\bar{\nu})} \left\{ \frac{|g_1|^2+|g_2|^2}{8} \right.
\\
&~~~~ + 2 \left( |g_3|^2 + |g_4|^2 \right) + |g_5|^2 + |g_6|^2 \\
&~~~~ - 8 e {\rm Re}\left[ A_R(2g_4^* +g_6^*)+A_L(2g_3^*+g_5^*) \right] \\
&~~~~ \left. + 64e^2 \left( \ln\frac{m_\mu}{m_e}-\frac{11}{8} \right)
\left( |A_L|^2+|A_R|^2 \right) \right\}.                
\end{split}
\end{align}
In our case, we obtain 
\begin{align}
	g_1&= g_2 = \frac{-i}{(4\pi)^2 m_{e_4}^2}H(x_\phi)f_{4e}^3 f_{4\mu}
	~, \\
	g_3&= g_4=g_5=g_5=\frac{-i}{4(4\pi)^2 m_{e_4}^2}A(x_\phi) f_{4e}^3 f_{4\mu}
	~,
\end{align}
where
\begin{align}
\nonumber
&
H(x_\phi) = \frac{2(1-x_\phi)+(1+x_\phi)\ln x_\phi}{(1-x_\phi)^3}
~,
\\
&
A(x_\phi) = \frac{x_\phi^2-2x_\phi\ln x_\phi -1}{(x_\phi-1)^3},       
\end{align}
with $x_\phi \equiv m_\phi^2 /m_{e_4}^2$.

\end{document}